\documentstyle[prl,twocolumn,aps,psfig,floats]{revtex}

\def\epsfig#1#2#3#4
         {
         \epsfysize=#2 \vbox{ \hglue#3 \epsfbox[#4]{#1} }
         }
\def\epsfigrot#1#2#3#4
         {
         \epsfxsize=#2
         \setbox\rotbox=\hbox to #2{\epsfbox[#4]{#1}}
         \vbox{\hglue#3 \rotl\rotbox}
         }
\newbox\rotbox
\input rotate

\begin{document}
\draft

\twocolumn[\hsize\textwidth\columnwidth\hsize\csname
@twocolumnfalse\endcsname

\title{Dense loops, supersymmetry, and Goldstone phases in two dimensions}

\author {J.L. Jacobsen,$\!^1$ N. Read,$\!^2$ and H. Saleur$\,^3$}

\address{$^1$ Laboratoire de Physique Th\'eorique et Mod\`eles Statistiques,
Universit\'e Paris-Sud, B\^atiment 100, F-91405 Orsay,
France\\$^2$ Department of Physics, Yale University, P.O. Box
208120, New Haven, CT 06520-8120, USA\\$^3$ Department of Physics
and Astronomy, University of Southern California, Los Angeles, CA
90089, USA}

\date{Date: \today}
\maketitle
\begin{abstract}
Loop models in two dimensions can be related to O($N$) models. The
low-temperature dense-loops phase of such a model, or of its
reformulation using a supergroup as symmetry, can have a Goldstone
broken-symmetry phase for $N<2$. We argue that this phase is
generic for $-2<\! N\! <2$ when crossings of loops are allowed,
and distinct from the model of non-crossing dense loops first
studied by Nienhuis {[\prl {\bf 49}, 1062 (1982)]}. Our arguments
are supported by our numerical results, and by a lattice model
solved exactly by Martins {\it et al.}{ [\prl {\bf 81}, 504
(1998)]}.
\end{abstract}

 \vspace{0.1in} ]

It has long been understood that theories of an $N$-component
scalar field $\phi$ with O($N$) symmetry in Euclidean space of
dimension $d$ can be related to statistical-mechanics models in
which configurations of loops are given Boltzmann weights
depending on their lengths and intersection properties
\cite{degennes}. In the $N\to 0$ limit, unwanted closed loops
vanish \cite{degennes}, and the usual self-avoiding walks
(polymers) are related to the generic critical point in the
O($N\to 0$) theory. The $N\to 0$ limit can be avoided by using a
model with a Lie superalgebra as symmetry (``supersymmetry'')
\cite{ParSou}. Other values of $N$ are also of interest. In $d=2$,
exact results based on lattice models \cite{nien} show that there
also exists a massless low-temperature (low-$T$) regime of the
model, for $-2<N\leq 2$. The properties of this phase---in which,
geometrically, the loops become dense---are well-understood using
Coulomb gas techniques \cite{nien}.

In this Letter, we revisit  the low-$T$ phase of the O($N \leq 2$)
models in $d=2$.  The supersymmetry approach, which can be
generalized to all integer $N$, makes it clear that for $N<2$
there should be a massless phase described by the Goldstone modes
of the broken O($N$) or supersymmetry, with very simple scaling
dimensions. We argue that the usual dense-loops phase \cite{nien},
in which loops never intersect, is not this Goldstone phase (in
contrast to earlier expectations \cite{oldhs}): in fact, it is not
generic, and an arbitrarily-weak perturbation that allows loop
intersections causes a cross-over to the generic Goldstone phase
for $-2<\! N\! <2$. We present numerical results that support our
interpretation. There is also a soluble model that appears to be
in the Goldstone phase \cite{Martins}.

The generic continuum O($N$)-invariant action $S$ for the scalar
field $\phi$ includes the interaction term
$-\lambda(\phi\cdot\phi)^2$ (by convention the Boltzmann weight is
$e^S$). In terms of loops, this is the model introduced by de
Gennes \cite{degennes}, in which the loops can cross, but for
$\lambda>0$ these crossings are disfavored. It has a second-order
transition for $d>2$, but for $d=2$, a transition occurs only for
$N\leq 2$; for $N>2$, there is only the high-$T$ or massive
unbroken-symmetry phase. (For $N$ not a positive integer, the
theory is supposed to be defined by analytic continuation of its
perturbation expansion. For $N$ integer, this can be made rigorous
by the use of supersymmetry, as we describe below.) When $-2
\leq\! N\!\leq 2$, the phase transition is second order, and the
associated critical exponents have been conjectured by Nienhuis
\cite{nien} and others \cite{sal86}. For $N<-2$, the transition is
expected to be first order.

The existence of the transition for $N\leq2$ implies the existence
of a low-$T$ phase. {}From the point of view of the generic O($N$)
$\phi^4$ theory, at low $T$ the symmetry is broken to O($N-1$)
according to Landau theory, and the low-$T$ phase would be
expected to be a Goldstone phase with massless excitations,
described by a nonlinear sigma model with target manifold
O($N$)/O($N-1$) $\cong S^{N-1}$, a sphere. For $N=2$, the O(2)
symmetry is not really broken, but there is a power-law phase with
continuously-varying exponents, the
Berezinskii-Kosterlitz-Thouless (BKT) phase \cite{bkt}. For $N<2$,
use of the perturbative beta function in the $S^{N-1}$ sigma
model, defined by analytic continuation in $N$, implies that the
coupling between the Goldstone modes in the sigma model becomes
weak at large length scales \cite{rs}, like Goldstone phases in
higher $d$. This yields a low-$T$ phase with simple forms for the
scaling dimensions. This is possible in $d=2$ for $N<2$ because
the Mermin-Wagner theorem no longer applies when the continuous
SO($N$) symmetry cannot be realized as unitary operations on a
vector field. We discuss this theory more fully below, using
supersymmetry.

In contrast, Refs.~\cite{dmns,nien} defined a particular lattice
mo\-del of strictly non-crossing loops.
This was done by truncating the high-$T$
expansion of a lattice version of the $\phi^4$ theory as above
\cite{dmns}, in order to simplify calculations of critical
properties while hopefully remaining in the same universality
class. The model has Boltzmann weight
\begin{equation}
e^S\equiv\prod_{\langle ij\rangle}
\left(1+K\vec{S}_i\!\cdot\!\vec{S}_j\right),\label{hexmod}
\end{equation}
where $i$, $j$ are vertices on the honeycomb graph, $\langle ij
\rangle$ denotes an edge of the graph, and the variables
$\vec{S}_i$ are $N$-component fixed-length spins, $\vec{S}_i^2=N$.
The partition function can be evaluated as $Z=\sum N^L K^E$ where
the sum is over graphs consisting of self-avoiding
mutually-avoiding loops, $L$ is the number of loops, $E$ the
number of edges they occupy, and $K\sim 1/T$. The critical $K$ for
this model is known exactly, $ K_{\rm
c}=\left(2+\sqrt{2-N}\right)^{-1/2}$ for $N\leq 2$ \cite{nien}.
The transition is second-order for $-2\leq\! N\!\leq 2$ only, and
the critical exponents at this critical point, which is known as
{\em dilute loops} (or as self-avoiding walks for $N\to 0$), are
known by a variety of techniques \cite{nien,sal86}. The low-$T$
region for each $-2<\!\!N\!\!<2$ is attracted to a unique massless
phase (that varies with $N$) for all $K_{\rm c}<\! K \! <\infty$.
This phase has nontrivial scaling dimensions, which are again
known \cite{nien,DupSal}. It has become known as {\em dense loops}
(or as dense polymers for $N\to 0$). In the limit $N\to 2$, this
phase coincides with the critical endpoint of the BKT phase (which
appears in the $N=2$ model for $K\geq K_{\rm c}$). For $N<-2$, we
expect that the dense-loops phase is massive.

Both the critical and dense-loops phases are believed to be
universal as continuum phases, in the sense that they are
independent of the choice of lattice used in defining them. Recent
work \cite{Kondev} on the $T=0$ limits of certain lattice models
identifies other massless phases of non-intersecting {\em
fully-packed} loops, that depend on the lattice used. We are not
concerned with these here, nor with the transition for $N>2$ in
the region $|K|>1/N$, where the Boltzmann weight (\ref{hexmod})
can be negative \cite{Wu}.

{}From the critical exponents for dilute loops, it is natural to
hope \cite{nien}, that they are in the same universality class as
the transition in the generic O($N$)
$\phi^4$ theory.
The idea is that any weak repulsion of the lines leads to a
crossover to the infinitely-repulsive limit in which the loops
never cross, so that the latter fixed point governs the transition
for all $\lambda>0$. We know of no reason to doubt this. As a
check, we can reintroduce intersections of the lines in the dilute
critical theory. The graphs of the expansion of the generic O($N$)
model that are not present in the truncated model would involve in
particular crossings of the loops, or multiple occupancies of
edges. In the continuum description, the most relevant (and most
generic) of these is where two lines cross. This may be included
as a perturbation of the dilute critical theory by adding to the
action the integral of a multiple of the corresponding scalar
operator, the so-called 4-leg operator, which has conformal weight
\cite{sal86}
$h_4=g-(g-1)^2/(4g)$,
where we parameterized $N=-2\cos\pi g$, $g\in [1,2]$. We see that
this operator is marginal at $N=2$ and irrelevant (i.e.~$h_4>1$)
for $-2\leq \!N\! <2$. This strongly suggests that allowing more
general interactions or configurations of loops in the lattice
model would not change the universality class, at least when the
perturbations are small, and hence that the {\em dilute} critical
point \cite{nien} is indeed generic.

The model of non-intersecting loops was obtained by truncating the
high-$T$ expansion, and this would be expected not to change the
behavior for sufficiently high $T$ (small $K$); apparently this
remains true down to $T=T_{\rm c}$. However, this leaves open the
possibility that the low-$T$ behavior of the models is different,
which we will now examine. The dense-loops phase is described by a
theory similar to that of the dilute case, with the same
parameterization of $N$ and $h_4$ as above, but with $g\in [0,1]$.
One sees that the 4-leg operator is now relevant in the whole
region $-2<\!N\!<2$.
As $N \searrow -2$, $h_4 \searrow -\infty$. Since the 4-leg
perturbation is relevant in the low-$T$ phase, it is {\em
dangerously} irrelevant in the critical theory.

It is thus clear that the {\em dense}-loops phase is non-generic
from an O($N$)-model point of view. In $d>2$, de Gennes' model
appears generic, as the topological effect of strict
non-intersection, present in $d=2$, is absent (though in $d=3$,
there remains the topology of knots and linking instead).
Even a model in $d>2$ of strict non-intersection, when
reduced to $d=2$ by confining $d-2$ dimensions to finite
intervals, gives an effective model in which intersections still
occur. There is also a symmetry interpretation for the
non-genericity: the dense phase of strictly non-intersecting loops
has a larger symmetry, U($N$) instead of O($N$), which is based on
the possibility of consistently orienting all the loops in the
partition sum \cite{rs,nien}. Though this symmetry may be broken
by the boundary conditions, the massless dense-loops phase appears
to be most easily understood in terms of a nonlinear sigma model
with this larger symmetry \cite{rs}. The 4-leg operator is a
symmetry-breaking perturbation, that reduces the symmetry to the
generic O($N$). It would be surprising if the large-length-scale
effect of this relevant symmetry-breaking perturbation led back to
the same higher-symmetry phase. We note that the importance
of the self-crossings has been observed in the  related
context of Lorentz lattice gases \cite{cao}.

We now argue that the flow induced by the 4-leg perturbation leads
to the generic Goldstone phase of the $\phi^4$ theory, and begin
by examining the structure of this phase, before giving numerical
evidence that such a flow does in fact occur.

The loop (high-$T$) expansion of the O($N$) model can be
reproduced by considering instead a model with OSp$(m|2n)$
symmetry, with $N=m-2n$; in particular, each closed loop incurs a
factor of $N$ \cite{ParSou,rs}. The model then makes complete
sense even for $N$ non-positive, (but $N$ must be an integer). The
phase diagram, beta functions, and scaling dimensions are the same
for all $m$ at fixed $N$ (though multiplicities of operators may
vanish for $m$ small \cite{rs}). The Goldstone phase should thus
be described by a sigma model with target space
OSp$(m|2n)$/OSp$(m-1|2n)\cong S^{m-1|2n}$, a supersphere, and a
single coupling constant $g_\sigma$ \cite{rs,salkauf}. The
perturbative beta function is, to leading order, $\beta\equiv
dg_\sigma/d\ln L\propto (N-2)g_\sigma^2$, so for $N<2$, the model
flows to weak coupling for $g_\sigma\geq 0$ (which is the expected
physical sign). Again, the Mermin-Wagner theorem does not apply
here, and symmetry-breaking is allowed for $N<2$. The
weak-coupling fixed point is a theory of free massless scalars,
with $m-1$ bosonic and $2n$ fermionic components, so is conformal
with central charge $c=m-1-2n=N-1$, as it would be for $S^{N-1}$.
There will be logarithmic corrections due to the
marginally-irrelevant coupling $g_\sigma$. We emphasize that in
the sigma model, OSp$(m|2n)$ symmetry does not allow any other
marginal or relevant couplings to be added to the action, so that
this fixed point is robust, unlike the dense-loops theory above.

Explicitly, the target manifold can be parameterized by commuting
coordinates $x_i$, $i=1,\ldots,m$, and anticommuting
coordinates $\eta_j$, $j=1,\ldots,2n$, subject to the
constraint $\sum_{i=1}^m x_i^2 +\sum_{j=1}^n
\eta_{2j-1}\eta_{2j}=1$. The action of the sigma model is
\begin{equation}
S=-{1\over g_\sigma}\int d^2r\left[\sum_{i=1}^m(\partial_\mu
x_i)^2+\sum_{j=1}^n\partial_\mu \eta_{2j-1}\partial_\mu
\eta_{2j}\right].
\end{equation}
The simplest example is $m=n=1$ ($N=-1$), which corresponds to the
$S^{0|2}$ supersphere. The constraint can be solved as
$x_1=1-{1\over 2}\eta_1\eta_2$. After elimination of $x_1$ and
some rescaling, one finds
\begin{equation}
S=-\int d^2 r \left[\partial_\mu\eta_1\partial_\mu\eta_2-
{g_\sigma\over 2}\eta_1\eta_2
\partial_\mu\eta_1\partial_\mu\eta_2\right].
\end{equation}
At long wavelengths the coupling $g_\sigma$ renormalizes towards
zero, and the theory becomes free symplectic fermions, with
central charge $c=-2$, as claimed. In general, for $N<1$, the
partition function of the fixed-point theory vanishes, $Z=0$.
Similar behavior is well-known in the dense-polymer ($N=0$) phase.

The supersymmetry analysis is valid for $N$ integer. For general
$N$, we are forced again to interpolate formally. The availability
of the supersymmetry results greatly strengthens our confidence in
this procedure and in the existence of the Goldstone phase for all
$N<2$.

Both at the dilute critical point and in the dense-loops phase,
the central charge is $c=1-6(g-1)^2/g$, in the parameterization
above. For dense loops, $g\in [0,1]$, this coincides with $c=N-1$
for $N=1$, $2$. The $N=2$ case is the BKT phase with $c=1$
throughout, and not of interest here. We claim that, for all
$-2<\!N\!<2$, the perturbed dense-loops phase flows to the
Goldstone phase, which is a distinct massless phase even when it
has the same $c$. For $1<\!N\!<2$, $c$ decreases during the flow,
but for $-2<\!N\!<1$ it increases. This latter behavior would not
violate Zamolodchikov's $c$-theorem, as the theories involved are
non-unitary. In particular, the $N=0$ ($c=-2$) dense-polymer
theory should flow to a $c=-1$ free theory. Note that for $N=1$
the partition function is $Z=1$ (after removing nonuniversal
constants) in both the dense-loops and Goldstone phases, a
consequence of $c=0$.

Some particularly interesting operators in loop models in general
are the $k$-leg operators (also known as fuseaux or watermelon
operators), insertions of which give the probability that $k$
lines terminate at the same point. In the general O($N$) point of
view, the leading such operator would be represented by $k$
insertions of the field $\phi$ at the same point (with no
derivatives), and must necessarily be in a totally-symmetric
rank-$k$ tensor representation of O($N$), which can be assumed to
be traceless on all pairs of indices. For the OSp$(m|2n)$
formulation, this becomes a supersymmetric tensor, that vanishes
when contracted using the same invariant bilinear form as in the
constraint in the sigma model. It is useful to choose the index
values of the set of tensors in a correlation function to occur in
distinct pairs, to force the lines in the loop model to connect
the positions of the operators in some specified way. This is
possible if $m$ is sufficiently large, a typical situation in
applications of supersymmetry.

At the dilute critical point and in the dense-loops phase, there
are nontrivial scaling dimensions $X_k$ (or conformal weights
$h_k=X_k/2$) for the $k$-leg operators, which are known exactly
\cite{sal86,DupSal} (for example, $h_4$ was given earlier).
(In these phases, the symmetric-tensor $k$-leg operators are in
general degenerate with others \cite{rs}, and in the dense-loops
phase, we assume there is a singlet 4-leg operator, which we used
as the perturbation above; it is the most relevant invariant
scalar operator.) In the Goldstone phases, on the other hand, the
symmetric tensors correspond to functions like spherical harmonics
on the target space, and their scaling dimensions tend to zero as
$g_\sigma\to 0$. Since the symmetry is broken for $N<2$, the
precise form of each operator depends on which components are
chosen. If the expectation of the field is $x_1=1$, all other
$x_i$ and $\eta_j=0$, then any component of the $k$-leg operator
can be rewritten in terms of the (rescaled, free) $m-1$ $x_i$'s
and the $2n$ $\eta_j$'s, and the correlation functions will
contain various logarithmic factors depending on which components
are chosen. It is clear that the scaling dimensions of the leading
$k$-leg operators are all zero, and there will be subleading
operators with integer conformal weights (the precise form of
correlation functions may require more attention to the limit
$g_\sigma\to 0$). When the dense-loops phase is perturbed by the
4-leg operator, it should be possible to see the effective scaling
dimensions of the $k$-leg operators (say, in two-point functions)
cross over to zero, for all $k$.

\begin{figure}[tbh]
\centerline{\psfig{figure=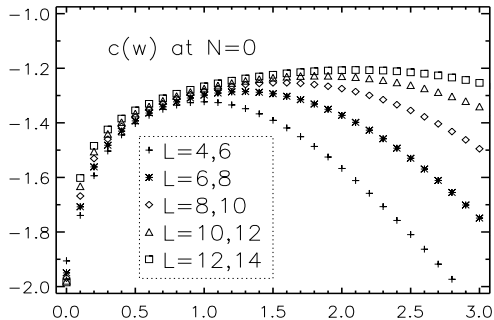,height=2.0in,width=3.0in,angle=0}}
\caption{Measures of the central charge vs.\ $w$ for $N=0$. The
conjectured value for $L\to\infty$ is $c=-1$ for all $w>0$.}
\label{fig:c}\vspace{-0.2in}
\end{figure}

Numerical transfer matrix calculations were done on a
square-lattice cylinder of circumference $L$. The edges of the
lattice are packed with one line per edge; two lines can meet at
each vertex in three ways, the weight being $1$ for a non-crossing
and $w$ for a crossing. The natural O($N$)-invariant boundary
condition for a periodic system is that the fields $\phi$ are
periodic, both for Bose and Fermi components. This implies that a
loop that winds the periodic direction is given the same factor
$N$ as a topologically trivial loop. In a conformally-invariant
system, $c$ can be extracted from the scaling of the free energy
per vertex, $f(L)=f(\infty)-\pi c/(6 L^2)$. We therefore use this
formula to extract the effective value of $c$ in a crossover,
using $f(L)$ at pairs of values of $L$. Results for $N=0$
(see Fig.\ \ref{fig:c}) show a clear trend away from
$c=-2$, and are consistent with a slow approach to $c=-1$
for $L\to\infty$.

\begin{figure}[tbh]
\centerline{\psfig{figure=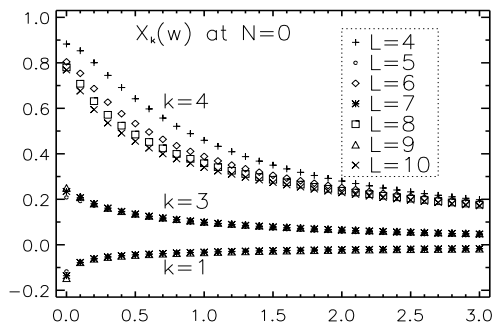,height=2.0in,width=3.0in,angle=0}}
\caption{Measures of the $k$-leg exponents $X_k=2h_k$ vs.\ $w$ for
$N=0$.
} \label{fig:h_k}\vspace{-0.1in}
\end{figure}

For the $k$-leg scaling dimensions, we use the lowest transfer
matrix eigenvalues $E_k$ in each sector with $k$ lines propagating
along the cylinder. From conformal invariance \cite{Cardy}, we
define effective values of $X_k$ at finite $L$ from
 $E_k-E_0=2\pi X_k/L$;
see Fig.\ \ref{fig:h_k} ($k=2$ is not included as, for $N=0$ in
our model, $E_2=E_0$ for all $L$ and $w$). The results are
consistent with $X_k=0$ in the fixed-point theory.  In a situation
with a marginally-irrelevant operator, one expects \cite{Cardy}
corrections to scaling to be $O(1/L\ln L)$, which presumably
explains the slow convergence.

\begin{figure}[tbh]
\centerline{\psfig{figure=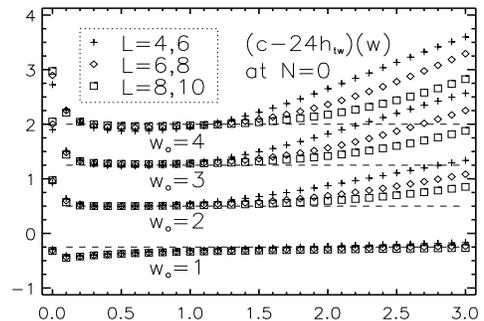,height=2.0in,width=3.0in,angle=0}}
\caption{Measures of $c-24h_{\rm tw}$ vs.\ $w$ for $N=0$. Dashed
lines show the prediction of Eq.\ (\ref{htw}).} \label{fig:ceff}\vspace{-0.1in}
\end{figure}

A useful check of our analysis is obtained by modifying the
boundary conditions, in a similar way as in Ref.\ \cite{rs}. If we
make the Fermi components of the $\phi$ fields obey antiperiodic
boundary conditions, this will give a non-contractible (winding)
loop the usual weight $w_{\rm e}=m-2n=N$ if its winding number is
even, and $w_{\rm o}=m+2n=N+4n$ if it is odd. In the Goldstone
phase, this will give the $n$ pairs of symplectic fermions the
antiperiodic boundary condition, while leaving the bosonic scalar
fields periodic. In the formula for $f(L)$, $c$ will be replaced
by $c-24h_{\rm tw}$, where $h_{\rm tw}$ is the lowest conformal
weight in the sector with boundary conditions twisted as
described. We expect $c-24h_{\rm tw}=m+n-1$, which can be
understood as the contribution of the twist operator of weight
$h=-1/8$ for each symplectic fermion pair.
Thus we expect
\begin{equation}
c-24h_{\rm tw}=N-1+3n={3w_o+w_e\over 4}-1,\label{htw}
\end{equation}
for general $N$, $n$. This appears to be confirmed numerically for
$N=w_{\rm e}=0$ and several values of $w_{\rm o}$ in Fig.\
\ref{fig:ceff}. Notice how this differs from the usual loop
gas behavior ($w=0$),
where $c-24h_{\rm tw}=1-6 \arccos^2(\frac{w_{\rm o}}{2})/(\pi^2 g)$
\cite{cardy00}.

For $N<-2$, we believe that the dense-loops phase is
massive, and hence flows to the generic Goldstone phase only when
$w$ is greater than a positive critical value.

Finally, it is important that a version of the
OSp$(m|2n)$-invariant model on the square lattice is integrable at
$w=(2-N)/4$ and can be solved by the Bethe ansatz \cite{Martins}.
The model is massless for $N<2$, and there is strong evidence that
the central charge is $c=N-1$, and that the lowest scaling
dimension, other than that of the identity operator, is zero; the
authors conjecture that these are the exact values, but do not
identify the phase with the Goldstone phase.

To conclude, we have argued that crossing of loops is a relevant
perturbation in the dense-loops low-$T$ phase for $-2<\!N\!<2$,
and that the long-distance behavior is governed by the Goldstone
phase instead, by analytical, numerical, and exact approaches.

This work was supported by the NSF under grant no.\ DMR-98-18259
(NR), and by the DOE (HS).

\end{document}